\documentclass[12pt]{article}
\usepackage{graphicx,amsmath}
\usepackage{graphicx,amsmath,lineno,color}

\usepackage{units}
\usepackage{caption} 
\usepackage{subcaption}
\usepackage[backend=biber,sorting=none,natbib=true,doi=true]{biblatex} 
\renewbibmacro{in:}{}
\usepackage[hidelinks,colorlinks=true,linkcolor=blue,urlcolor=blue,citecolor=blue]{hyperref}
\hypersetup{colorlinks=true}

\newlength{\dinwidth}
\newlength{\dinmargin}
\def\lapproxeq{\lower .7ex\hbox{$\;\stackrel{\textstyle                                                    
<}{\sim}\;$}}                                                    
\def\gapproxeq{\lower .7ex\hbox{$\;\stackrel{\textstyle                                                    
>}{\sim}\;$}}                                                    
\def\be{\begin{equation}}                                                    
\def\ee{\end{equation}}                                                    
\def\bea{\begin{eqnarray}}                                                    
\def\eea{\end{eqnarray}}

\def\sh{\hat s}
\def\sh2{{\hat s}^2}

\begin{document}                                                    
\titlepage                                                    
\begin{flushright}                                                    
\today \\                                                    
\end{flushright} 
\vspace*{0.5cm}

\begin{center}                                                    
{\Large \bf Oscillations in elastic scattering at large momentum transfer at the LHC?
\\}

\vspace*{1cm}
Per~Grafstr\"om  \\                       
\vspace*{0.5cm}                                                    
 Universit\`a di Bologna, Dipartimento di Fisica, 40126 Bologna, Italy\\
                        
\vspace*{1cm}                                                    
 
\begin{abstract}
The available data on elastic scattering at the LHC  at large momentum transfer $t$  in the  range $0.05 <|t|<1.0~\mathrm{GeV}^{2}$ have been analyzed in terms of possible oscillating structures. A clearly significant structure is seen in the data from the TOTEM collaboration at 13 TeV. Data measured by the same collaboration at 2.76~TeV, 7~TeV and 8~TeV are not statistically significant to confirm or reject the observation at~13 TeV. More data are needed to understand if the effect is real or an experimental artefact   \end{abstract}

\end{center}

\vspace{1cm}

\section{Introduction}
The possibile existence  of fine structures of the differential elastic cross section has  been discussed from time to time over the last 50 years. The possibility of oscillations in the very forward direction was discussed in  a paper of  Auberson, Kinoshita and Martin (AKM)~\cite{PhysRevD.3.3185}. In this paper it was shown, using axiomatic field theory  and   assuming unequal particle-particle and particle-antiparticle cross section asymptotically, that the scattering amplitude must have  infinitely many zeros in the forward direction. Such properties of the scattering amplitude might lead to oscillations of the differential cross section in the  forward direction. Hints of this type of oscillation were seen in the UA4/2 data~\cite{AUGIER1993448} and reported in Ref.~\cite{GAURON1997305}. 
No sign of those so-called  
 AKM oscillations was seen at the LHC  in the data from  ATLAS~\cite{ATLAS:2022mgx}  and TOTEM~\cite{TOTEM:2017sdy} analysed at very small $|t|$ i.e. $|t|<0.01~\mathrm{GeV}^{2}$ ~\cite{grafström2023akm}.

Oscillations in a much wider $t$-range and with wavelength order of magnitudes larger than the typical AKM oscillations have also been discussed since long.

Some hints of small amplitude deviations in terms of oscillating structures  were observed at the ISR~\cite{BARBIELLINI1972663},\cite{WHITE197323} in elastic proton-proton scattering. Similar indications were also seen in a Daresbury experiment  for elastic electron-proton scattering at an incident energy of 1.6 and 2.0 GeV \cite{BOTTERILL1973125}. Anomalous structures seen in the data at the SPS-collider and the Tevatron were furthermore discussed in Ref.~\cite{Barshay_1994}. In general the problems with the data sets in question are the relatively low statistics and it is difficult to draw any firm conclusions.

The ATLAS and TOTEM experiments at the LHC have studied elastic scattering at centre-of-mass energies ranging from 2.76 TeV to 13 TeV giving a  new opportunity to seek for  anomalous structure in terms of  possible small oscillating  deviations from the leading structure. Oleg Selyugin has published a series of papers in this respect and has a very comprehensive approach. Some examples can be found in  Refs.~\cite{Selyugin_2019},\cite{Selyugin_2021},\cite{selyugin2022anomalies} and \cite{selyugin2023new}. In a very recent  paper (see Ref.~\cite{selyugin2023new2}) Selyugin  looks for those anomalous  effects doing a common $\chi^2$-fit to all the data available at the LHC and as well a number of data sets from the SPS-collider and the Tevatron. Moreover, he uses the data from the smallest $t$-values up to the largest available values. He concludes that a non standard term of oscillating nature is  indeed present in the scattering amplitude over a wide energy range.

In this note we have a slightly different approach and more focused on a reduced data set. We will use the only very high statistics data set available at this point, i.e. the data from the TOTEM experiment at 13~TeV.  More than $ 10^9 $ events were collected here with an integrated luminosity well above $100~ \mathrm{nb}^{-1}$. ATLAS also has a similar data set but the results are still being analysed and have not yet been published. The other data sets at the LHC have at least one order of magnitude lower statistics.  We will use  the very high statistics data of TOTEM at 13~TeV as a starting point and determine the parameters characterising a possible oscillating addition to the elastic scattering amplitude. In a second step we will then see to what extent this amplitude is compatible with the other data sets at the LHC covering the centre-of-mass energies from 2.76~TeV to 8~TeV.  In this investigation we are not interested in the short wavelength and the very small $t$-region which characterise possible AKM-oscillations, which we discussed in Ref.~\cite{grafström2023akm}. Instead  we have chosen data sets in the $t$-range $0.05 <|t|<1.0~\mathrm{GeV}^{2}$ which  exclude the Coulomb and Coulomb-nuclear interference region and extend beyond the dip and bump structure.   

\section{Phenomenological description of the data}

In order to see if the differential elastic cross section exhibits some fine structure in terms of oscillating deviations from the leading amplitude it is important to find a model that gives a smooth description of the main shape of the cross section. In this note we  use the most simple model 
 possible describing the differential elastic cross section  in the region of the dip and the bump but  outside the Coulomb and Coulomb-nuclear interference region. The simple proposal  of Phillips and Barger \cite{PHILLIPS1973412}, gives a surprisingly good description of the elastic data at the LHC. The amplitude is just taken  as the sum of two exponentials with a relative phase in between them.
\begin{equation}
\label{nuclamp}
\frac{d\sigma_{el}}{dt}=|f_{pb}|^{2}=|\sqrt{A}e^{Bt/2}+\sqrt{C}e^{(Dt/2 +i\Phi)}|^2
\end{equation}
Measurements by TOTEM~\cite{2015527} and ATLAS~\cite{ATLAS:2022mgx} have clearly shown that in the energy range of the LHC  
and in the $t$-range $0.05 < |t| < 0.3~\mathrm{GeV}^2 $ it is necessary to add to the simple exponent a quadratic
and cubic term in the exponent to get a good description. 
Thus the first term in the simple Phillips-Barger formula of Eq.~\ref{nuclamp} has been modified according to this.

In order to try to describe a possible  oscillatory pattern  we use the ansatz of Selyugin that define an oscillatory term like:
\begin{equation}
\label{nuclamposc}
f_{osc}(t)= ih_{osc}(1+i)\cdot \ln\hat{s}\cdot J_{1}(\tau)/ \tau.
\end{equation}
$J_{1}$ is the Bessel function of the first order and the  $\hat{s}$ is defined as $\hat{s}=s~e^{-i\pi/2}$ with $s$ being the centre-of-mass energy squared normalized to $s_{0} =1~\mathrm{GeV}^{2}$.
$\tau$ is a scaling variable defined as 
\begin{equation}
\label{tau}
\tau=\pi (\phi_{0}-t )/t_{0},
\end{equation}
The parameter~$\phi_{0}$ is set to 0 and thus remain two additional parameters, i.e $h_{osc}$ and $t_{0}$ to be determined in a  fit to the experimental data
\section {TOTEM   data at \texorpdfstring{$\sqrt s$}{Lg}=13 TeV. }

\begin{figure}[ht]
\begin{subfigure}{0.49\textwidth}
\includegraphics[width=0.9\linewidth, height=6cm]{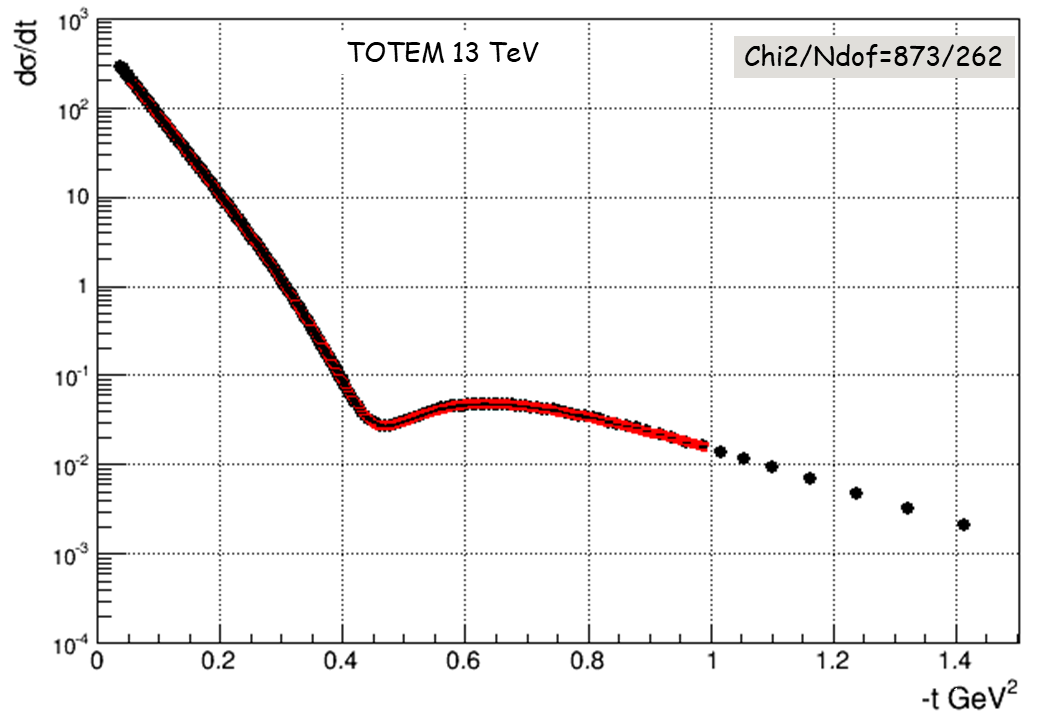} 
\caption{}
\label{fo1}
\end{subfigure}
\begin{subfigure}{0.49\textwidth}
\includegraphics[width=0.9\linewidth, height=6cm]{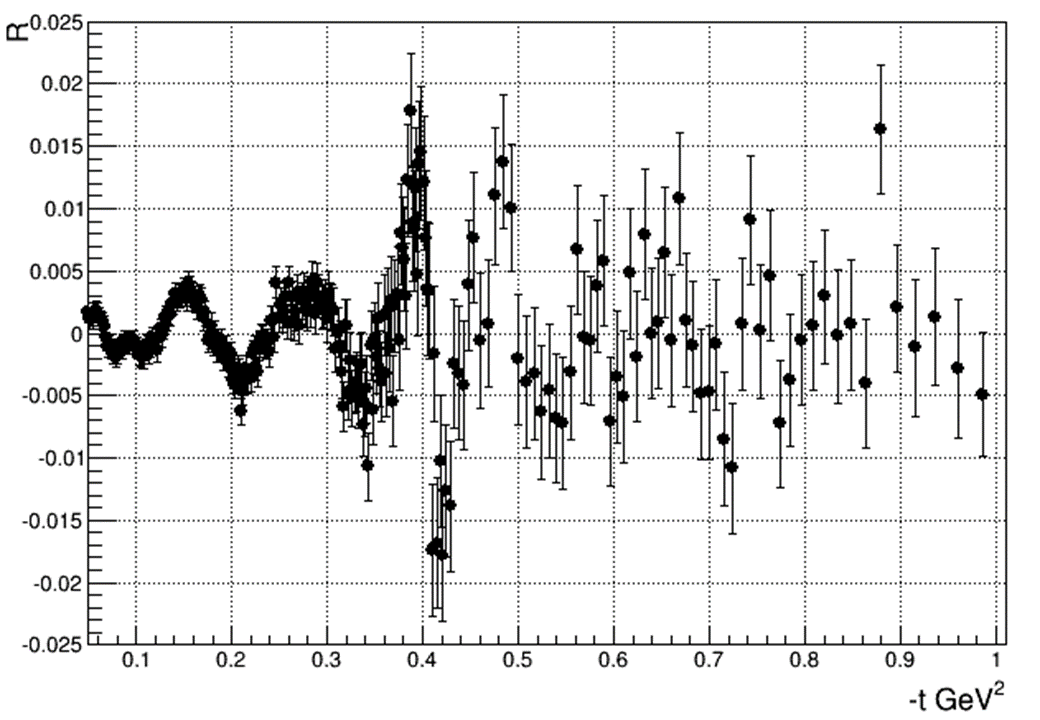}
\caption{ }
\label{fo2}
\end{subfigure}
    \caption{ (a) TOTEM data at 13 TeV~\cite{refId0,} as a function of $-t$. The data have been fitted with the Philipps-Barger model~\cite{PHILLIPS1973412} using Eq.~\ref{nuclamp}. (b) The ratio $R$  defined in the text and in  Eq.~\ref{defR} as a function of $-t$. }
\end{figure}

The result of the fit to the TOTEM data at 13~TeV~\cite{refId0} using Eq.~\ref{nuclamp}, i.e. without the oscillating term $f_{osc}$ is shown in
 Figure~\ref{fo1}. By eye the fit looks very good but the $\chi^{2}/\mathrm{Ndof}   =3.33$ which of course is very bad. The explanation to the bad $\chi^{2}/\mathrm{Ndof}$ value is seen in Figure~\ref{fo2} where the ratio R is plotted as a function of $t$ in the fitted range.  The ratio R is defined as :
\begin{equation}
\label{defR}
R=\frac{\frac{d\sigma_{el}}{dt}(\mathrm{data})}{\frac{d\sigma_{el}}{dt}(\mathrm{pbfit})}-1.
\end{equation}
where \(\frac{d\sigma_{el}}{dt}(\mathrm{data})  \) is the elastic differential cross section as measured by TOTEM at 13~TeV and \(\frac{d\sigma_{el}}{dt}(\mathrm{pbfit}) \) represents the differential cross section using the Philipps-Barger model and the fitted values of the parameters. A very pronounced oscillating deviation from the fit result at the level of about 1 $\%$ is seen. There is no doubt that these deviations are statistically significant but it should also be said that the uncertainties used for the TOTEM data are only the statistical uncertainties. What is seen in Figure~\ref{fo2} can of course be an experimental effect, i.e. a subtle detector effect not understood or it can also be an artifact of the unfolding method used to calculate the differential elastic cross section from raw data. However, at this point we assume that the effect is real. 

\begin{figure}[htb]
\begin{subfigure}{0.49\textwidth}
\includegraphics[width=0.9\linewidth, height=6cm]{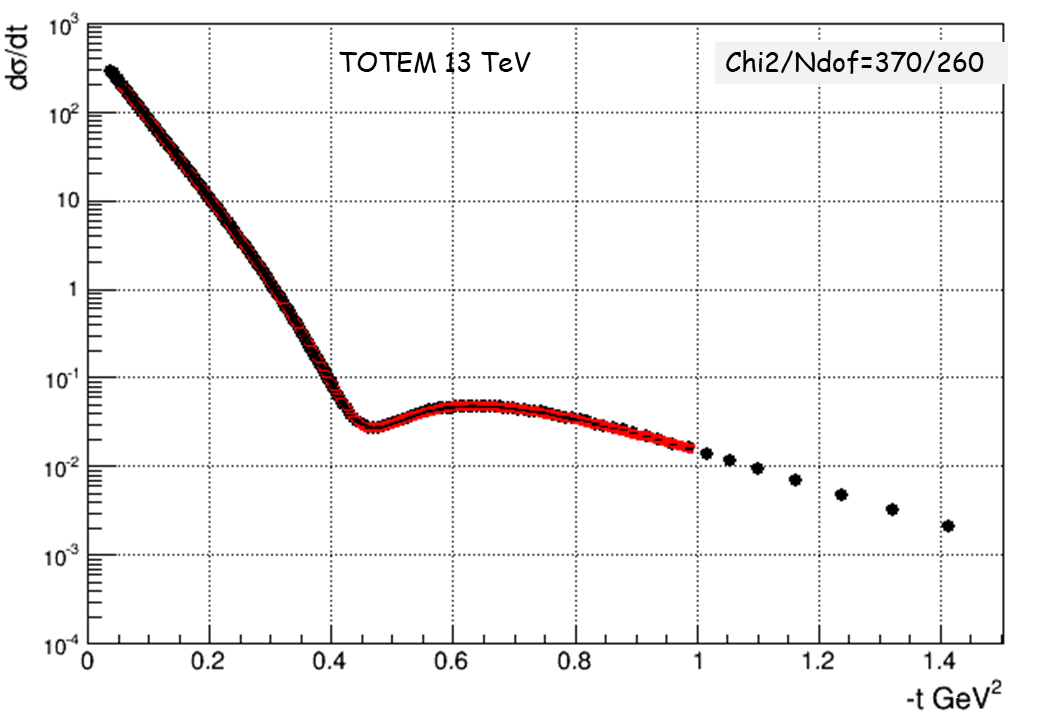} 
\caption{ }
\label{fo3}
\end{subfigure}
\begin{subfigure}{0.49\textwidth}
\includegraphics[width=0.9\linewidth, height=6cm]{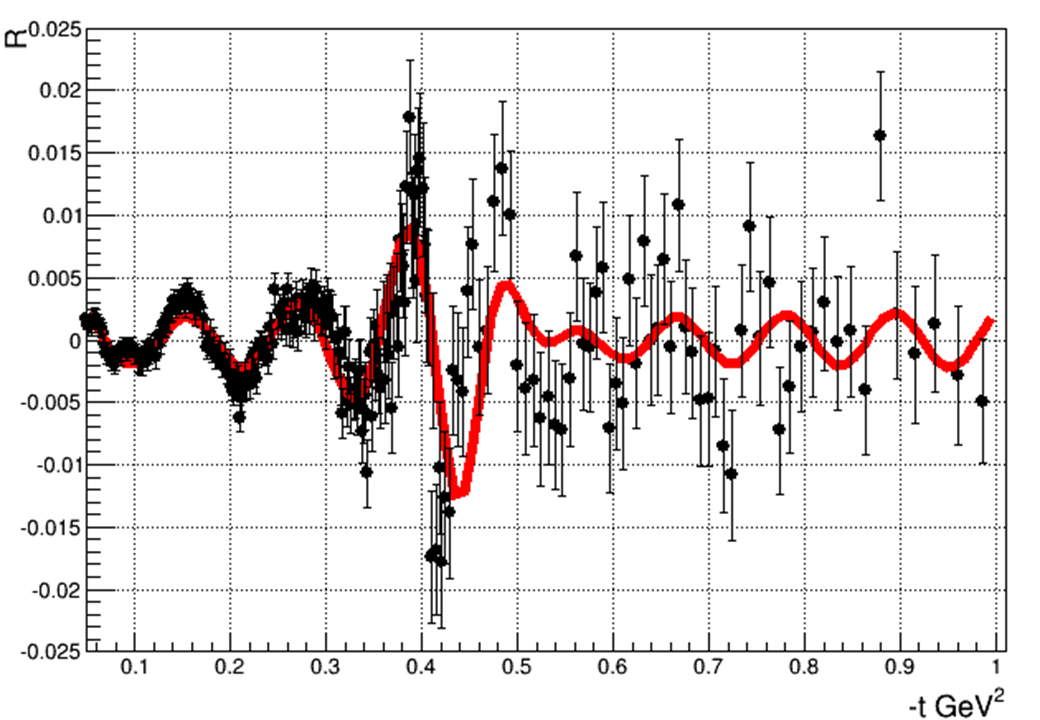}
\caption{ }
\label{fo4}
\end{subfigure}
\caption{ (a) TOTEM data at 13 TeV~\cite{refId0,} as a function of $-t$. The data have been fitted with the Philipps-Barger model~\cite{PHILLIPS1973412}  using Eq.~\ref{nuclamp} and adding an oscillatory term as given by Eq.~\ref{nuclamposc}. (b)    The ratio $R$  defined in the text and in Eq.~\ref{defR} as a function of $-t$. The red curve represents the oscillatory contribution found in the fit. }
\end{figure}

The TOTEM data have been refitted a second time and instead of using the amplitude $f_{pb}$ the amplitude $f_{pb}+f_{osc}$ has been used. The parameters found in first the fit using only $f_{pb}$ have been used thus leaving only two free parameters in the second fit, i.e. $h_{osc}$ and $t_{0}$. The result is shown in Figure \ref{fo3}. No evident difference between the first and second fit is seen by eye but the $\chi^{2}/\mathrm{Ndof}$ is improved from 3.33 to 1.42. As can be seen in Figure~\ref{fo4} the ansatz of Eq.~\ref{nuclamposc} gives a surprisingly good description of the oscillating pattern. 

\begin{figure}[ht]
\begin{subfigure}{0.49\textwidth}
\includegraphics[width=0.9\linewidth, height=6cm]{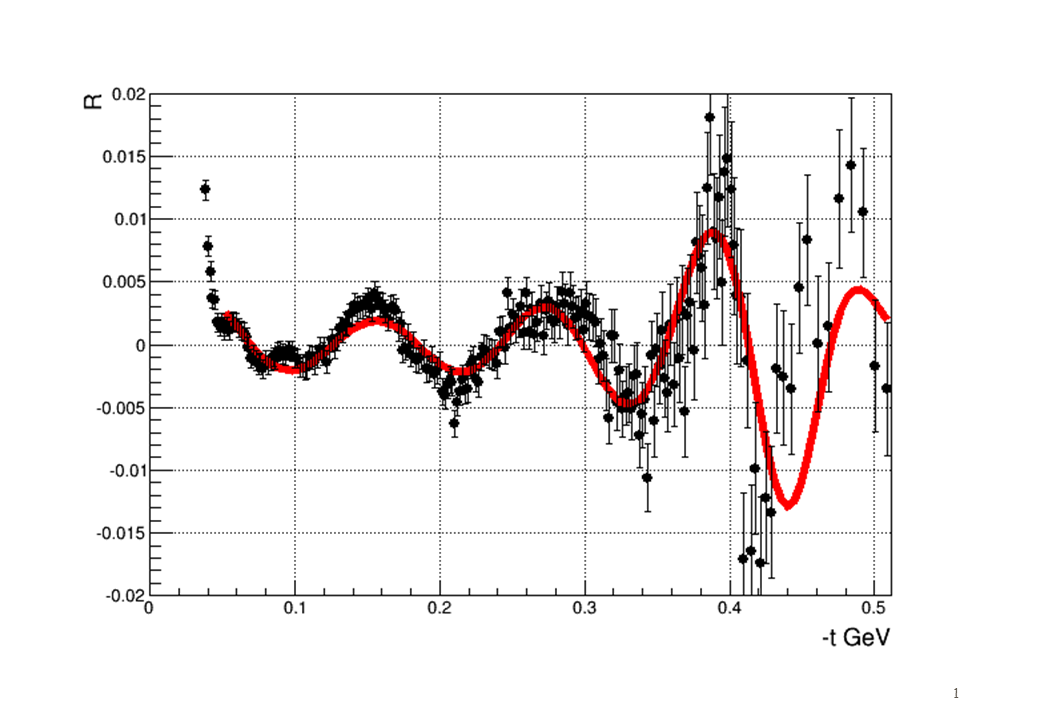} 
\caption{}
\label{fo11}
\end{subfigure}
\begin{subfigure}{0.49\textwidth}
\includegraphics[width=0.9\linewidth, height=6cm]{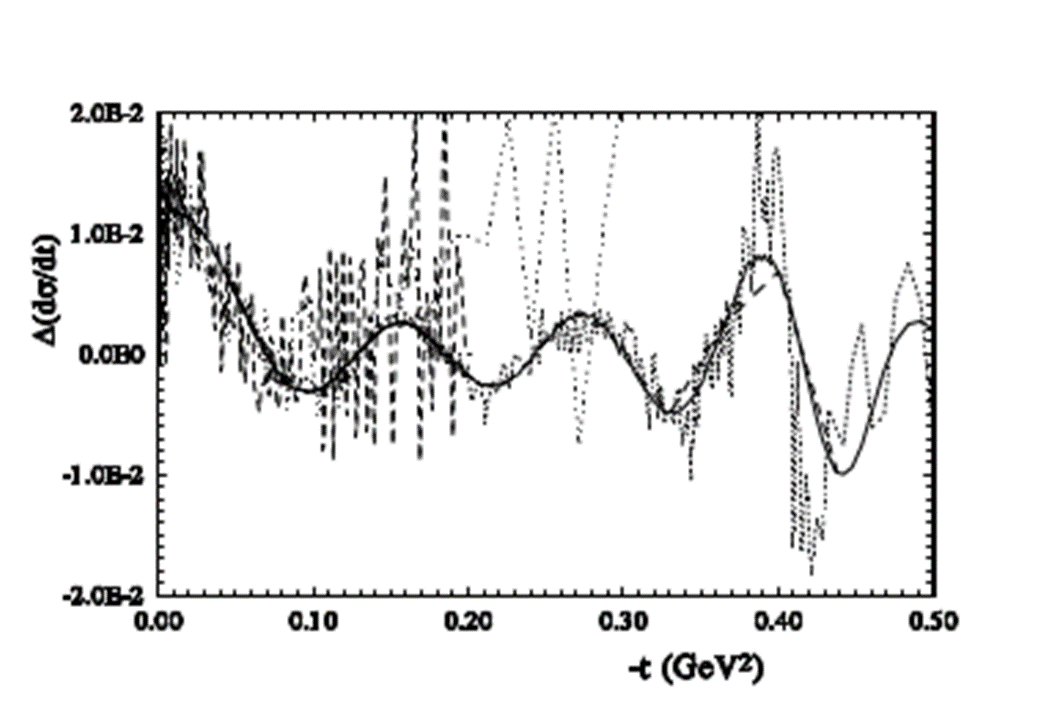}
\caption{ }
\label{fo12}
\end{subfigure}
    \caption{ (a) The oscillating pattern obtained using the Philipps-Barger model\cite{PHILLIPS1973412} and using the oscillating term given by Eq.~\ref{nuclamposc}.    (b) The oscillating pattern obtained in Ref.~\cite{selyugin2023new2} by using the HEGS model and the oscillating term given by Eq.~\ref{nuclamposc}. The $t$-range in the two figures is given by the $t$-range used in the Figure 4 of Ref.~\cite{selyugin2023new2}. }
\end{figure}

It should also be said that the more sophisticated analysis of Selyugin  in  Ref.~\cite{selyugin2023new2} using the full machinery of the HEGS model gives a picture very close to our result seen in Figure~\ref{fo4}. This is illustrated in Figure~\ref{fo11} and in~\ref{fo12} where the fitted oscillating pattern from this note is compared to the result of Selyugin. There is a very good agreement between the fitted curve for the two approaches.
 
If this effect is a real effect and not an instrumental effect or not an   artefact of the unfolding methods used in the analysis, a consistent picture  ought to emerge looking at the other data sets at the LHC as well.

\clearpage
\section {TOTEM data at 2.76 TeV, 7 TeV and 8 TeV}
\begin{figure}[b]
\begin{subfigure}{0.49\textwidth}
\includegraphics[width=0.9\linewidth, height=6cm]{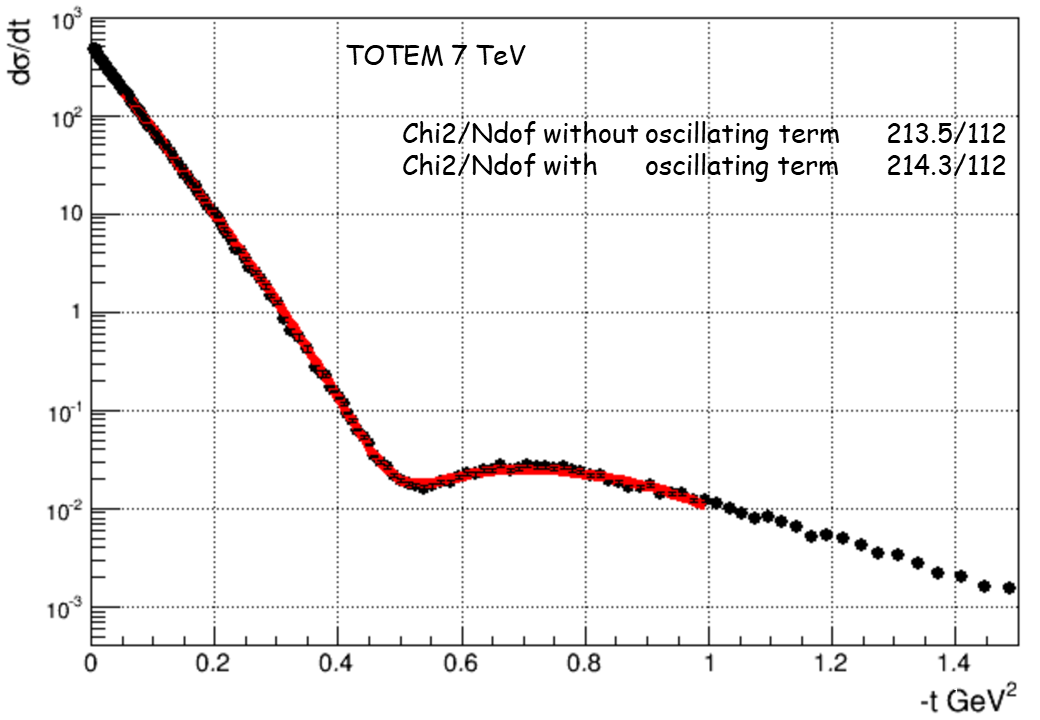} 
\caption{ }
\label{fo5}
\end{subfigure}
\begin{subfigure}{0.49\textwidth}
\includegraphics[width=0.9\linewidth, height=6cm]{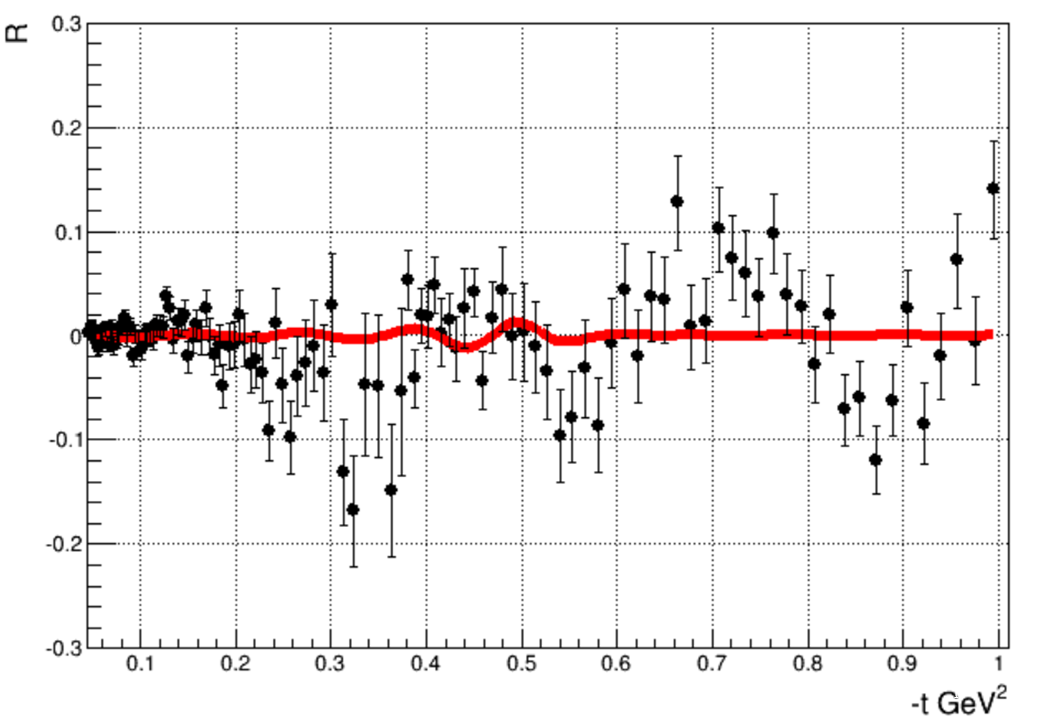}
\caption{ }
\label{fo6}
\end{subfigure}
\caption{ (a) TOTEM data at 7~TeV~\cite{Antchev_2013} as a function of $-t$. The data have been fitted with the Philipps-Barger model~\cite{PHILLIPS1973412} using Eq.~\ref{nuclamp}  with and without adding an oscillatory term as given by Eq.~\ref{nuclamposc}. The parameters of the oscillating term have been taken from the parameters determined at 13~TeV and applying $\ln s$ scaling. (b) The ratio $R$  defined in the text and in Eq.~\ref{defR} as a function of $-t$. The red curve represents the oscillatory contribution as determined at 13~TeV and applying $\ln s$ scaling. For more details see the corresponding text.}
\end{figure}

\begin{figure}[b]
\begin{subfigure}{0.49\textwidth}
\includegraphics[width=0.9\linewidth, height=6cm]{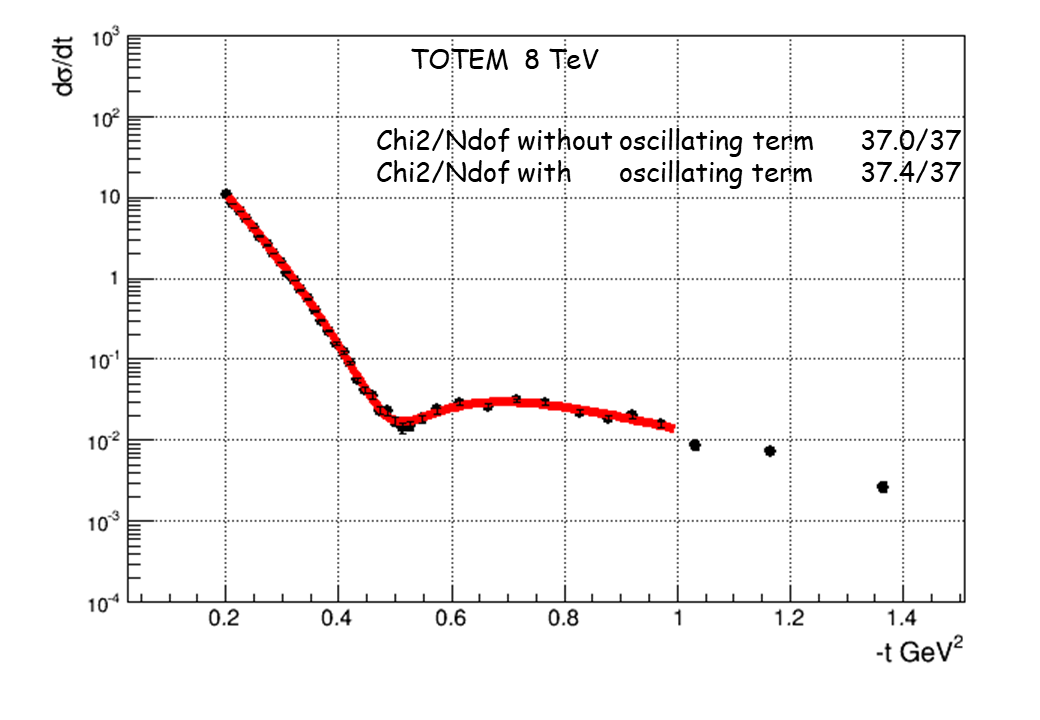} 
\caption{ }
\label{fo7}
\end{subfigure}
\begin{subfigure}{0.49\textwidth}
\includegraphics[width=0.9\linewidth, height=6cm]{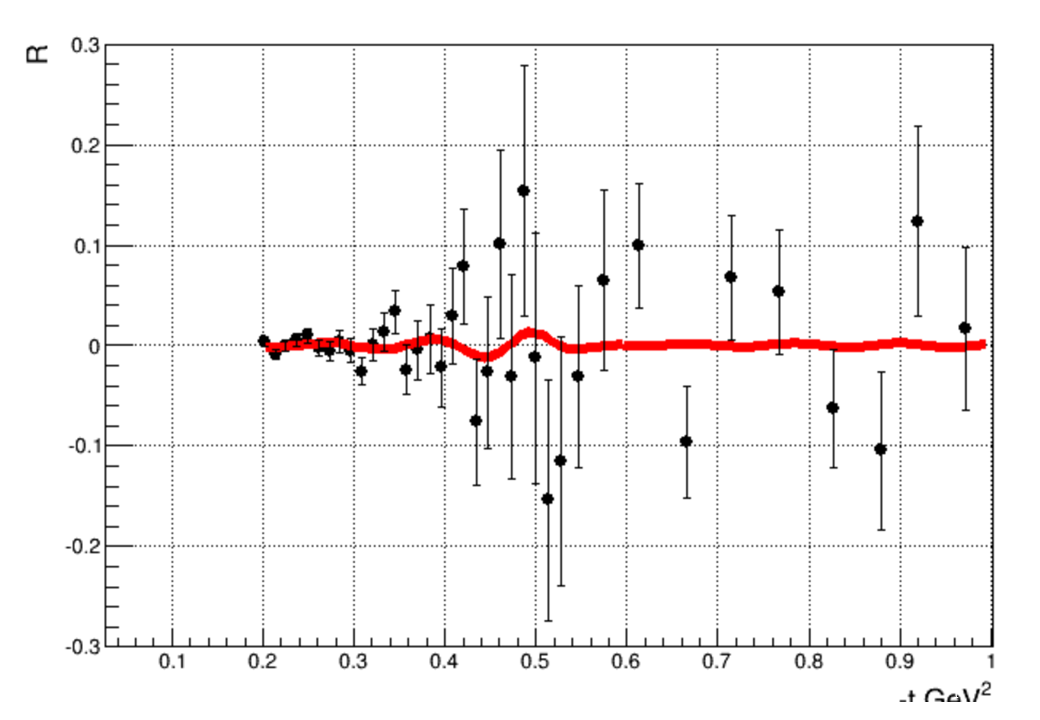}
\caption{ }
\label{fo8}
\end{subfigure}
\caption{ (a) TOTEM data at 8~TeV~\cite{thetotemcollaboration2021characterisation} as a function of $-t$. The data have been fitted with the Philipps-Barger model~\cite{PHILLIPS1973412} using Eq.~\ref{nuclamp}  with and without adding an oscillatory term as given by Eq.~\ref{nuclamposc}. The parameters of the oscillating term have been taken from the parameters determined at 13~TeV and applying $\ln s$ scaling. (b) The ratio $R$ defined in the text and in Eq.~\ref{defR} as a function of $-t$. The red curve represents the oscillatory contribution as determined at 13~TeV and applying $\ln s$ scaling. For more details see the corresponding text.}
\end{figure}

\begin{figure}[ht]
\begin{subfigure}{0.49\textwidth}
\includegraphics[width=0.9\linewidth, height=6cm]{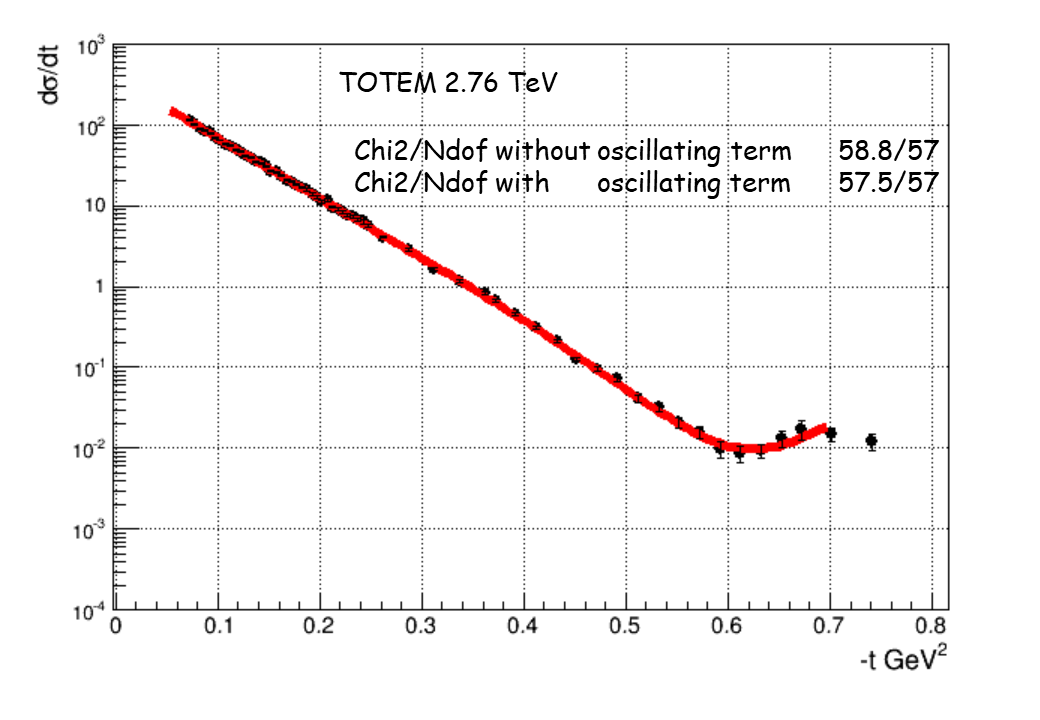} 
\caption{ $TOTEM$}
\label{fo9}
\end{subfigure}
\begin{subfigure}{0.49\textwidth}
\includegraphics[width=0.9\linewidth, height=6cm]{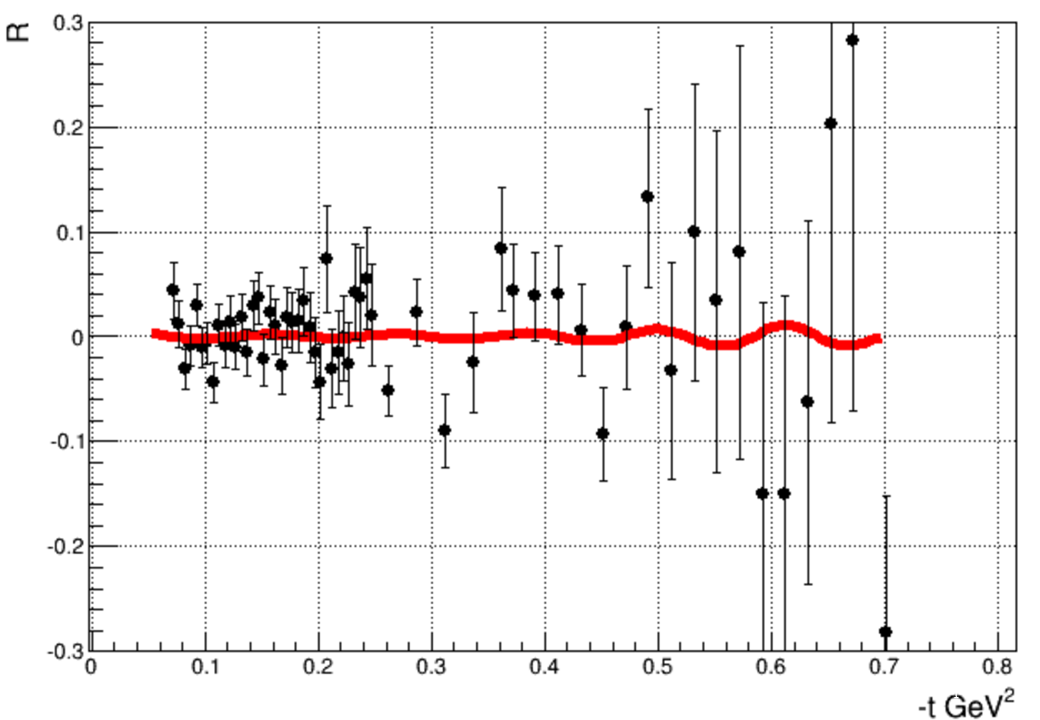}
\caption{$ATLAS$ }
\label{fo10}
\end{subfigure}
\caption{ (a) TOTEM data at 2.76 TeV~\cite{Antchev_2020} as a function of $-t$.The data have been fitted with the Philipps-Barger model~\cite{PHILLIPS1973412}  using Eq.~\ref{nuclamp} with and without adding an oscillatory term as given by Eq.~\ref{nuclamposc}. The parameters of the oscillating term have been taken from the parameters determined at 13~TeV and applying $\ln s$ scaling. (b) The ratio $R$  defined in the text and in Eq.~\ref{defR} as a function of $-t$. The red curve represents the oscillatory contribution as determined at 13~TeV and applying $\ln s$ scaling. For more details see the corresponding text.} 
\end{figure}
There are three additional data sets from the TOTEM experiment that go beyond the dip-bump structure. Unfortunately the statistics in those data sets are more than an  order of magnitude lower than in the case of the 13~TeV data set and thus  the statistically fluctuations are significantly larger. This very significant difference in  statistics between the 13~TeV data set and the other three data sets is clearly seen in  the  fluctuations of the ratio $R$ defined by Eq.~\ref{defR}. Below will be shown the ratio $R$ for the three additional data sets and the scale has been changed from ~$\pm2.5 ~\%$ at 13~TeV  to~$\pm30 ~\%$ in order to give a clear picture of the fluctuations.

In Figure~\ref{fo5} and in Figure~\ref{fo6}  are shown  the results  using the 7~TeV data set~\cite{Antchev_2013} which  has most statistics of the three remaining data sets. The high $t$ data is taken with an integrated luminosity of $6~\mathrm{nb}^{-1}$. In Figure~\ref{fo5} the Philipps-Barger amplitude given by Eq.~\ref{nuclamp}  has been used with and without adding the oscillator term. The $\chi^{2}/\mathrm{Ndof}$ is not very good but is not improved by adding  the oscillatory term.  The oscillatory term has been taken as Eq.~\ref{nuclamposc} using the values of $h_{osc}$ and $t_{0}$ found in the fit to the high statistics data at 13~TeV  and the $\ln s$ scaling has been applied. A $\ln s$ scaling of $t_{0}$ has also been tried but there is no significant change in the result. It should be said that the $\ln s$ scaling between 13~TeV and 7~TeV is of little importance with a ratio of $\ln s$ equal 1.07. From Figure~\ref{fo6} is it clear that the statistics is not enough to confirm or to reject the oscillator pattern at the level of $1\% $ found in the 13~TeV data. 

The data sets at 8~TeV~\cite{thetotemcollaboration2021characterisation} and 2.76~TeV~\cite{Antchev_2020} have been investigated in the same manner as the 7~TeV data set, i.e using the parameters of the oscillating term found at 13~TeV and applying the $\ln s$ scaling. These data sets have even less statistics than the 7~TeV data set with integrated luminosities of order of a $\mathrm{nb}^{-1}$ The results are shown in Figures~\ref{fo7}, \ref{fo8} and Figures~\ref{fo9}, \ref{fo10} respectively and as can be seen also here the statistics is too low for any conclusion.

\section{Discussion}
 In this note we have shown that the high statistics differential elastic cross section as measured by the TOTEM experiment at 13~TeV has a clearly significant oscillating component. The $\chi^{2}/\mathrm{Ndof}$
 takes the value of 873/262 in a fit without an oscillating term and adding an oscillating term the $\chi^{2}/\mathrm{Ndof}$ changes to 370/260. This clearly confirms the observation of Selyugin in Ref.~\cite{selyugin2023new2}. However the data sets of TOTEM at the energies 2.76~TeV, 7~TeV and 8~TeV do not have enough statistics to confirm or reject the oscillating pattern found at 13~TeV. 

  Selyugin in Ref.~\cite{selyugin2023new2} uses 25 data sets for a common fit  in the energy interval from $\sqrt{s}=500$ GeV up to  $\sqrt{s}=13$ TeV and sees a  significant difference depending on including or not including an oscillation term.  May be this result is mainly driven by the very high statistics data set at 13~TeV giving a dominating contribution to the global $\chi^{2}$?
 
 It is clear that the structure seen at 13~TeV has to be explained. Is it an experimental effect or is it some interesting physics behind? One candidate for an experimental effect could be oscillations generated by the unfolding method. If this is the case or not can only be tested by the TOTEM authors.
 If it is a real effect it should also be seen in the ATLAS very high statistics data set where the analysis is ongoing.

\vspace{\baselineskip} 
{\bf\Large Acknowledgements}

Many thanks to Hugh Montgomery  who made us aware of previous discussions of oscillations related to ISR and Daresbury data and who has given several very useful comments to this note. 
Also many thanks to  Valery Khoze and Misha Ryskin for reading and  commenting this note  as well as to Hasko Stenzel for comments and  for improving the typesetting.  In addition I am grateful to the members of the ATLAS-ALFA group with whom I have discussed this topic many times.
\vspace{\baselineskip} 
\clearpage
\printbibliography 

\end{document}